\documentstyle[aps]{revtex}
\begin{document}
\title{Role of lepton flavor violating (LFV) muon decay in Seesaw model\ and LSND}
\author{M.Jamil Aslam and Riazuddin}
\address{National Centre for Physics, Quaid-i-Azam University,\\
Islamabad, Pakistan.}
\maketitle

\begin{abstract}
The aim of the work is to study LFV in a newly proposed Seesaw model of
neutrino mass and to see whether it could explain LSND excess. The
motivation of this Seesaw model was that there was no new physics beyond the
TeV scale. By studying $\mu \rightarrow 3e$ in this model, it is shown that
the upper bound on the branching ratio requires Higgs mass $m_{h}$ of a new
scalar doublet with lepton number $L=-1$ needed in the model has to be about 
$9$ TeV. The predicted branching ratio for $\mu \rightarrow e\nu _{l}\bar{\nu
}_{l}$ is too small to explain the LSND.

\end{abstract}

\section{Introduction}

The problem of physics beyond the Standard model (SM) has been studied for a
considerable length of time. In the past few years some progress has been
made to understand this new physics, among them LFV is the most promising
candidate. At present, we have rigorous bounds on LFV $\mu $ decay,e.g.\cite
{j1} 
\begin{equation}
{\cal B}\left( \mu \rightarrow 3e\right) \leq 10^{-12}  \label{1}
\end{equation}
Using experimental bounds on these three body decays the corrosponding
bounds on two body decays are calculated in \cite{ja1,j2}, 
\begin{equation}
{\cal B}\left( Z\rightarrow \mu e\right) \leq 1.7\times 10^{-13}  \label{2}
\end{equation}
and \cite{j2} 
\begin{eqnarray}
{\cal B}\left( J/\psi \rightarrow \mu e\right) &\leq &4\times 10^{-13}
\label{3} \\
{\cal B}\left( \Upsilon \rightarrow \mu e\right) &\leq &2\times 10^{-9}
\label{4} \\
{\cal B}\left( \Phi \rightarrow \mu e\right) &\leq &4\times 10^{-17}
\label{5}
\end{eqnarray}

At present, the best experimental limit on the branching ratio of $
Z\rightarrow \mu e$ decay is (95\%C.L.) 
\begin{equation}
{\cal B}\left( Z\rightarrow \mu e\right) \leq 1\cdot 7\times 10^{-6}.
\label{6}
\end{equation}
The possible source of the suppression of the bounds found in Eqs. (\ref{2}
)-(\ref{5}) are discussed in \cite{ja1,j2,j3,j4}.

The most alluring issue in the present day physics is whether or not the
neutrinos have non zero mass. In the minimal standard model of particle
interactions, neutrinos are massless.To generate a small neutrino mass in
this model, there is an effective dimension five operator 
\begin{equation}
{\cal L}_{eff}=\frac{f_{ij}}{\Lambda }L_{i}L_{j}\Phi \Phi  \label{7}
\end{equation}
where $L_{i}=\left( \nu _{i},l_{i}\right) _{L}$ is the usual left-handed
lepton doublet, $\Phi =\left( \phi ^{+},\phi ^{0}\right) $ is the usual
scalar Higgs doublet, and $\Lambda $ is an effective large mass scale\cite
{j5}. This operator has different tree-level realizations: (I) the canonical
seesaw mechanism with right handed neutrino\cite{j6}; (II) the model having
Higgs triplet\cite{j7}; and (III) the model having heavy Majorana ferimon
triplet\cite{j8}. These new interactions exist at higher mass scale. In the
usual Seesaw mechanism, in order to have a very small mass for left handed
neutrino, the corresponding mass for the right handed neutrino has to be
very large, i.e. of order $10^{3}$ TeV. Recently a new Seesaw model of
neutrino mass is proposed with the motivation that there is no new physics
beyond the TeV scale\cite{j9}. In this model the smallness of mass for right
handed neutrino does not require a very heavy right handed neutrino. This
mechanism requires $m_{N}\sim 1$TeV. However in this model, a new Higgs
doublet $\eta $ with lepton number $-1$ is also necessary. The right handed
neutrino $N$ and Higgs $\eta $ can give rise to LFV processes. We identify
an effective operator in Standard\ Model and show that the scale of new
physics $\Lambda $ must be $\Lambda \geq 5$ TeV. In the Ma's model, $\mu
\rightarrow 3e$ can proceed through $N$ and $\eta $ exchange at loop level.
Using experimental limit on this process, the box diagrams provide the most
stringent limit on mass of Higgs $\eta $ ($m_{h}\geq 9$ TeV). We also show
by constructing an effective $Z\rightarrow \mu e$ vertex and its realization
through $\eta $ and $N$ exchange that no limit is put on $m_{h}$.

The experimental evidence of the neutrino masses comes, from three anomalous
effects; LSND\ excess\cite{j10,j11}, atmospheric neutrino anomaly\cite
{j12,j13,j14} and the solar neutrino deficit\cite{j15,j16,j17,j18}.

In addition to the three neutrinos, a sterile neutrino is needed to explain
the three effects in terms of the neutrino flavor oscillations. But still
the problem is unresolved \cite{j19}. Attempts to explain all the data in
terms of the three massive neutrinos is excluded by the latest data\cite{j20}
.

Atmospheric anomaly and the solar neutrino deficit can be explained in terms
of neutrino flavor oscillations, but LSND\ excess can not be explained on
these lines, because of its small transition probability ($P_{\bar{\nu}_{\mu
}\rightarrow \bar{\nu}_{e}}=\left( 2.5\pm 0.6\pm 0.4\right) \times 10^{-3})$
) \cite{j21}. It either requires a sterile neutrino or some mechanism other
than neutrino flavor oscillation and LFV\ is one of the candidates\cite{j22}
. We analyze the consquences of small LFV interactions to explain LSND\
excess and show that the branching ratio for $\mu ^{+}\rightarrow e^{+}\nu
_{l}\bar{\nu}_{l}$ turns out to be too small to explain the LSND excess.

\section{LFV in Seesaw model of neutrino mass and bounds on new Higgs Meason
mass}

A new Seesaw model of neutrino mass has been proposed\cite{j9}, where right
handed fermion singlets $N_{i}$ with lepton number $L=0$ are added to the
minimal SM together with a second scalar doublt $(\eta ^{+},\eta ^{0})$ with
lepton number $L=-1$. The fermion singlet is allowed to have Majorana mass $
m_{N}$ with the effective interaction $f_{ij}\bar{N}_{iR}\left( \nu
_{jL}\eta ^{0}-l_{jL}\eta ^{+}\right) $. The smallness of Seesaw neutrino
mass $(m_{\nu }=\frac{m_{D}^{2}}{m_{N}})$ can be explained by a rather small
value of $m_{N}$, if $m_{D}$ comes from $\left\langle \eta ^{0}\right\rangle 
$ instead of $\left\langle \phi ^{0}\right\rangle $ because $\left\langle
\eta ^{0}\right\rangle <<\left\langle \phi ^{0}\right\rangle $, and is of
the order $1$ TeV and as such can be observed experimentlly. As the
motivativation of the model is that there is no new physics beyond the TeV
scale, therefore masses of the new Higgs scalar doublet, which are necessary
in this model, should not be larger than a few TeV. We study this question
visa viz the experimental bounds on $\mu \rightarrow 3e$ and $Z\rightarrow
\mu e$.

First we note that an effective operator invariant under the symmetries of
the Standard Model that can induce the decay $\mu \rightarrow 3e$ is 
\begin{equation}
\frac{1}{\Lambda ^{4}}\left( \bar{L}_{\mu }e_{R}\Phi \right) \left( \bar{L}
_{e}e_{R}\Phi \right)  \label{8}
\end{equation}
which is of dimension $8$ and here $\Phi $ is the Standard Model Higgs. On
inserting the vacuum expectation value $\left\langle \phi ^{0}\right\rangle
=v$ it would generate the four-fermion interaction 
\begin{equation}
\frac{v^{2}}{\Lambda ^{4}}\left( \bar{\mu}_{L}e_{R}\right) \left(
e_{L}e_{R}\right) .  \label{9}
\end{equation}
This operator leads to $\mu \rightarrow 3e$ with the branching ratio 
\begin{equation}
{\cal B}\left( \mu \rightarrow 3e\right) =\left( \frac{v^{2}}{4\sqrt{2}
G_{F}\Lambda ^{4}}\right) ^{2}  \label{10}
\end{equation}
Inserting $v=174$ GeV, the branching ratio (\ref{10}) gives $\Lambda \geq
4.7 $ TeV. However, the above effective operator can not be induced in the
Ma's model at tree level. In this model $\mu \rightarrow 3e$ occurs in the
one loop shown in Fig. 1. The couplings at each vertex can be taken from\cite
{j9}.

The amplitude can be written as follows\cite{j23}: 
\begin{eqnarray}
iT\left( \mu \rightarrow 3e\right) &=&2\sum_{i,j}\left( f_{\mu
i}^{*}f_{ei}f_{ej}^{*}f_{ej}\right) \int \frac{d^{4}k}{\left( 2\pi \right)
^{4}}\bar{v}(p_{1})\not{k}\left( \frac{1-\gamma _{5}}{2}\right) v(p_{2}) 
\nonumber \\
&&\times \bar{u}(p_{3})\not{k}\left( \frac{1-\gamma _{5}}{2}\right)
v(p_{4})\left[ \frac{1}{k^{2}-m_{h}^{2}}\right] ^{2}\left[ \frac{1}{
k^{2}-m_{i}^{2}}\right] \left[ \frac{1}{k^{2}-m_{j}^{2}}\right]  \label{11}
\end{eqnarray}
It is assumed that the loop momenta is very high, i.e. $k\rightarrow \infty $
so that the external momenta are neglected. After calculating the loop
integration Eq. (\ref{11}) becomes [we take $m_{i}=m_{j}=m$]

\begin{equation}
T\left( \mu \rightarrow 3e\right) =\frac{1}{\left( 4\pi \right) ^{2}\times
4m_{h}^{2}}\times \sum_{i,j}\xi _{i}\xi _{j}\left\{ \bar{v}(p_{1})\gamma
^{\alpha }\left( 1-\gamma _{5}\right) v(p_{2})\bar{u}(p_{3})\gamma _{\alpha
}\left( 1-\gamma _{5}\right) v(p_{4})\right\} A\left( x\right)  \label{12}
\end{equation}

where, $x=\left( \frac{m^{2}}{m_{h}^{2}}\right) $ and $f_{\mu
i}^{*}f_{ei}=\xi _{i}$, $f_{ej}^{*}f_{ej}=\xi _{j}$, giving the decay width 
\begin{equation}
\Gamma \left( \mu \rightarrow 3e\right) =\left[ \frac{1}{\left( 4\pi \right)
^{2}\times 4m_{h}^{2}}\right] ^{2}2\sum_{i,k}\xi _{i}\xi
_{k}^{*}\sum_{j,l}\xi _{j}\xi _{l}^{*}A\left( x\right) A\left( x\right)
\times \frac{m_{\mu }^{5}}{192\times \pi ^{3}},  \label{13}
\end{equation}
where,

\begin{equation}
A\left( x\right) =\left\{ \frac{1-x^{2}+x\ln \left( x^{2}\right) }{2\left(
x-1\right) ^{3}}\right\} .  \label{14}
\end{equation}
This gives the branching ratio 
\begin{equation}
{\cal B}\left( \mu \rightarrow 3e\right) =\left[ \frac{1}{4\left( 4\pi
\right) ^{2}}\right] ^{2}\left( \frac{2}{G_{F}^{2}}\right) \sum_{i,k}\xi
_{i}\xi _{k}^{*}\sum_{j,l}\xi _{j}\xi _{l}^{*}\left( \frac{x^{2}}{m^{4}}
\right) [A\left( x\right) ]^{2}.  \label{15}
\end{equation}
Assuming that all the Yukawa couplings are of the order unity and taking $
m=1 $ TeV the experimental bound (\ref{1}) is obtained for $x=1.12\times
10^{-2}$ giving $m_{h}=9$ TeV.

We now consider $Z\rightarrow \mu e$ for which the effective operator is 
\begin{equation}
g_{Z\mu e}\bar{L}_{\mu }\gamma ^{\mu }D_{\mu }L_{\mu }  \label{16}
\end{equation}
or 
\begin{equation}
\tilde{g}_{Z\mu e}\bar{\mu}_{R}\gamma ^{\mu }D_{\mu }e_{R}  \label{17}
\end{equation}
which are renormalizable operators of dimension $4$ so that the effective
coupling constants are dimensionless. $D_{\mu }$ is the covariant
derivative. The effective operator (\ref{16}) is induced in the present
model by renormalizable interaction represented by the Feynman diagrams
shown in Fig. 2. Each of the above diagrams is logaritmically divergent; but
this divergence cancels in the sum if we note that the first diagram
involves $Z$- lepton coupling in the form 
\begin{eqnarray}
&&\left( g_{V}+g_{A}\right) \frac{g_{2}}{2\cos \theta _{w}}  \nonumber \\
&=&\left\{ -\frac{1}{2}\left( 1-4\sin ^{2}\theta _{w}\right) +\left( -\frac{1
}{2}\right) \right\} \frac{g_{2}}{2\cos \theta _{w}}  \nonumber \\
&=&-\left( 1-2\sin ^{2}\theta _{w}\right) \frac{g_{2}}{2\cos \theta _{w}}
\label{18}
\end{eqnarray}
while in the second diagram $Z\rightarrow \eta \bar{\eta}$ gauge coupling is 
$\frac{g_{2}}{2\cos \theta _{w}}\left( 1-2\sin ^{2}\theta _{w}\right) $. In
fact the two diagrams exactly cancel in the limit $m_{Z}=0$ and leptons mass$
=0$. Using the dimensional regularization these diagrams together give 
\begin{equation}
iT=\frac{g_{2}\left( 1-2\sin ^{2}\theta _{w}\right) }{2\cos \theta _{w}}
\frac{f^{2}}{\left( 4\pi \right) ^{2}}\varepsilon _{\mu }\bar{u}
(k_{1})\gamma ^{\mu }\left( 1-\gamma _{5}\right) v(k)\left( \frac{m_{Z}^{2}}{
m_{h}^{2}}\right) I(x)  \label{19}
\end{equation}
where with $x=m_{i}^{2}/m_{h}^{2}$ and where we have kept only term linear
in $m_{Z}^{2}/m_{h}^{2}$; $I(x)$ is given by 
\begin{equation}
I(x)=-\frac{1}{36\left( 1-x\right) }\left\{ 2-\frac{3x}{1-x}+\frac{6x^{2}}{
\left( 1-x\right) ^{2}}+\frac{6x^{3}}{\left( 1-x\right) ^{3}}\ln x\right\} .
\label{20}
\end{equation}
In obtaining the final result we have neglected a convergent contribution
from the second diagram which is proportional to the lepton mass. Using $
\sin ^{2}\theta _{w}\simeq \frac{1}{4}$, Eq. (\ref{19}) gives 
\begin{equation}
\Gamma \left( Z\rightarrow \mu e\right) =\frac{G_{F}}{\sqrt{2}}\frac{%
m_{Z}^{3}}{6\pi }\frac{1}{2}\left( \frac{1}{4\pi }\right) ^{2}f^{4}\left( 
\frac{m_{Z}^{2}}{m_{h}^{2}}\right) ^{2}I^{2}  \label{21}
\end{equation}
while 
\begin{equation}
\Gamma _{Z}^{\text{tot}}=8\frac{G_{F}}{\sqrt{2}}\frac{m_{Z}^{3}}{6\pi }.
\label{22}
\end{equation}
Thus the branching ratio is given by 
\begin{eqnarray}
{\cal B}\left( Z\rightarrow \mu e\right) &=&\frac{1}{16}f^{4}\left( \frac{1}{
16\pi ^{2}}\right) ^{2}\left( \frac{m_{Z}^{2}}{m_{h}^{2}}\right) ^{2}I^{2} 
\nonumber \\
&=&2.5\times 10^{-6}\left( \frac{m_{Z}^{2}}{m_{h}^{2}}\right) ^{2}I^{2}
\label{23}
\end{eqnarray}
where we have taken the Yukawa couplings $f\simeq 1$. Taking the two extreme
limits $x\rightarrow 1$ and $x\rightarrow 0$, $I(x)$ is respectively $\frac{1
}{24}$ and $\frac{1}{18}$; the branching ratio bound (\ref{2}) can be
satisfied for $m_{h}\geq m_{i}$ $\simeq 500$ GeV. Thus no limit is put on $
m_{h}$.

\section{The anomalous muon decay and LSND}

The standard muon decay $\mu ^{+}\rightarrow e^{+}\nu _{e}\bar{\nu}_{\mu }$
has no $\bar{\nu}_{e}$ which is found at LSND. Instead of the neutrino
flavor oscillation this excess can be found through the LFV muon decay $\mu
^{+}\rightarrow e^{+}\nu _{l}\bar{\nu}_{l}$. The anomalous muon decay $\mu
^{+}\rightarrow e^{+}\nu _{l}\bar{\nu}_{l}$ (where $l=e$, $\mu $, or $\tau$
) can occur via $Z$-exchange as shown in Fig. 3, where the effective
Lagrangian $Z\mu e$ vertex can be written as\cite{j2} 
\begin{equation}
{\cal L}_{eff}=g_{Z_{\mu e}}\bar{\mu}_{L}\gamma ^{\alpha }e_{L}Z^{\alpha
}+h.c.  \label{24}
\end{equation}
Here, $g_{Z_{\mu e}}$ is the effective coupling of the LFV vertex which is
constrainted from the experimental bound on the branching ratio of $\mu
\rightarrow 3e$. The coupling through the diagram of Fig. 3, contributes to
the $A\left( \mu \rightarrow e\nu _{l}\bar{\nu}_{l}\right) $ amplitude a term

\begin{equation}
A\left( \mu \rightarrow e\nu _{l}\bar{\nu}_{l}\right) =\frac{g_{Z_{\mu
e}}g_{Z\nu \bar{\nu}}}{M_{Z}^{2}-s}\left[ \bar{v}(p_{1})\gamma ^{\alpha
}v(p_{2})\bar{u}(p_{4})\gamma _{\alpha }\left( 1-\gamma _{5}\right)
v(p_{3})\right]  \label{25}
\end{equation}
The corresponding decay width becomes 
\begin{eqnarray}
\Gamma \left( \mu \rightarrow e\nu _{l}\bar{\nu}_{l}\right) &=&\left( \frac{
g_{Z_{\mu e}}g_{Z\nu \bar{\nu}}}{M_{Z}^{2}}\right) ^{2}\frac{m_{\mu }^{5}}{
192\pi ^{3}}2.  \nonumber \\
&=&g_{Z_{\mu e}}^{2}\left( \frac{G_{F}}{\sqrt{2}}\right) ^{2}\frac{m_{\mu
}^{5}}{192\pi ^{3}}  \label{26}
\end{eqnarray}
where we have used that $g_{Z\nu \bar{\nu}}=\left( \frac{g_{2}}{2\cos \theta
_{w}}\right) \frac{1}{2}$. This gives the branching ratio

\begin{eqnarray}
{\cal B}\left( \mu \rightarrow e\nu _{l}\bar{\nu}_{l}\right) _{Z-exch.}
&=&g_{Z_{\mu e}}^{2}  \nonumber \\
&=&\left( 4\sqrt{2}G_{F}m_{Z}^{2}\right) .{\cal B}\left( Z\rightarrow \mu
e\right)  \nonumber \\
&=&\left( 0.55\right) .{\cal B}\left( Z\rightarrow \mu e\right)  \label{27}
\end{eqnarray}
where we have used Eq. (\ref{22}). If we use the bound (\ref{2}), we obtain 
\begin{equation}
{\cal B}\left( \mu \rightarrow e\nu _{l}\bar{\nu}_{l}\right) _{Z-exch.}\leq
10^{-13}.  \label{28}
\end{equation}
which is much too small compared to the needed branching ratio for $\mu
\rightarrow e\nu _{l}\bar{\nu}_{l}$ implied by 
\begin{equation}
P_{\bar{\nu}_{\mu }\rightarrow \bar{\nu}_{e}}=\left( 2.5\pm 0.6\pm
0.4\right) \times 10^{-3}.  \label{29}
\end{equation}
Even if one uses the direct limit (\ref{6}) it still remains small $\simeq
10^{-6}$.

Since there is no $\nu _{R}$ in the Standard Model, one cannot write an
operator of the form (\ref{8}) with $e_{R}$ replaced by $\nu _{R}$. One can
however write two dimension $9$ operators as in\cite{j25} which would
generate $\Delta L=2$ decay $\mu ^{+}\rightarrow e^{+}\bar{\nu}_{e}\bar{\nu}
_{l}$ with a branching ratio which could explain LSND excess with out any
conflict with $\Delta L=0$ processes like $\mu \rightarrow 3e$ with a scale
of new physics at a rather low value $\Lambda \simeq 360$ GeV. However, such
operators cannot be induced in the present model at tree level. If one
considers the box diagrams for $\mu \rightarrow e\nu _{l}\bar{\nu}_{l}$,
which gives the same result as in Eq. (\ref{15}), namely 
\begin{equation}
{\cal B}\left( \mu \rightarrow e\nu _{l}\bar{\nu}_{l}\right) =\left( \frac{1
}{64\pi ^{2}\times m_{i}^{2}}\right) ^{2}\frac{2}{G_{F}^{2}}\left(
81f^{8}\right) \left[ xA(x)\right] ^{2}
\end{equation}
which gives the branching ratio to be $\approx 10^{-12}$ for $f\approx 1$
and $m_{h}\approx 9$ TeV as previously found. This is much too small
compared to the required value $\simeq 2.5\times 10^{-3}$.

\section{Conclusion}

By studying process $\mu \rightarrow 3e$ at the loop level, we have put a
bound on the mass of the new Higgs boson $m_{h}$ needed in the Seesaw model
of neutrino masses \cite{j9}. Taking the Yukawa-couplings $f$ to be order $1$
and mass of the heavy neutrino to be $1$ TeV as in \cite{j9}, we found the
bound on $m_{h}\geq 9$ TeV. No limit is put on $m_{h}$ from the experimental
bound on $Z\rightarrow \mu e$. The LSND neutrino anomaly in terms of the
decay $\mu ^{+}\rightarrow e^{+}\nu _{l}\bar{\nu}_{l}$ which requires the
branching ratio of the new decay to be about $\left( 1.5-3\right) \times
10^{-3}$ can not be explained in the present model. Lastly we wish to
discuss the senstivity of the above bound on $f$. First of all we note that
atmospheric anomaly required $m_{\nu }\geq 5\times 10^{-2}$ eV which in Ma's
model implies that $f^{2}\geq 5\times 10^{-2}$ keeping $m_{i}\simeq 1$ TeV.
Then one obtains no bound on $m_{h}.$

{\bf Acknowledgments}

One of us (Jamil) like to thank Professor Fayyazuddin for stimulating
discussions and Gilani for ever ready help. This work was supported by a
grant from Pakistan Council of Science and Technology. A part of work of
(Jamil) was also supported by Mumtaz Riazuddin fellowship.

{\bf Figure Captions:}

1. Box diagrams for $\mu \rightarrow 3e$

2. One loop diagrams for $Z\rightarrow \mu e$

3. Anamolous muon decay

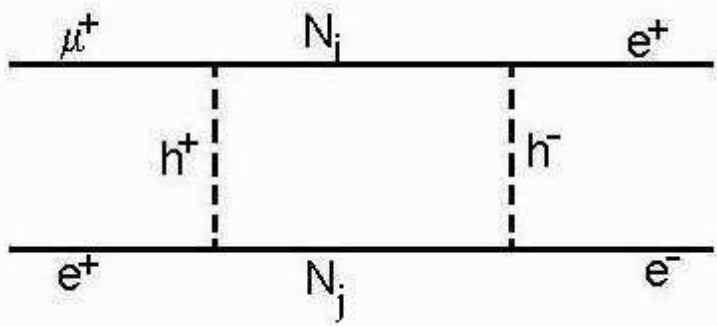
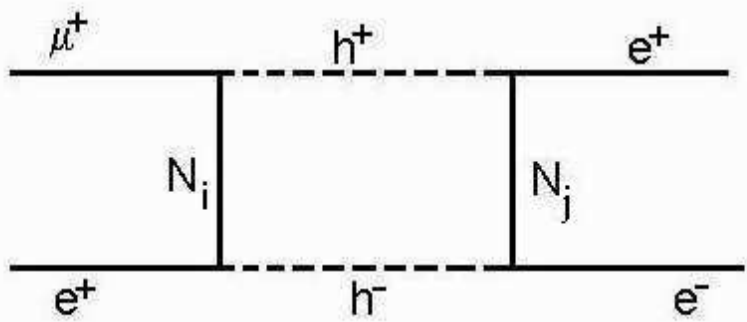
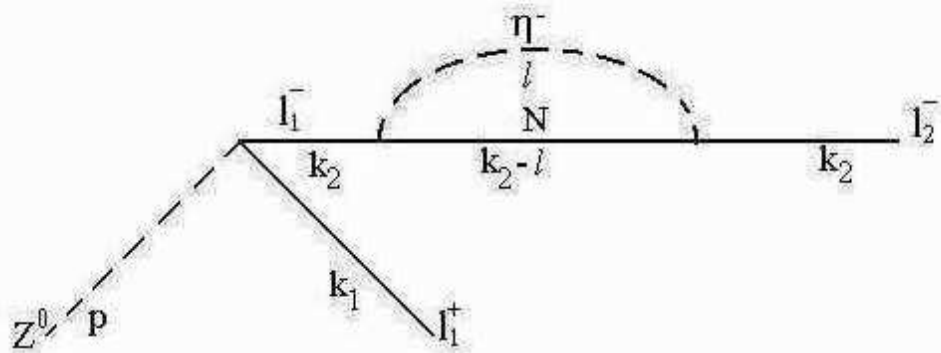
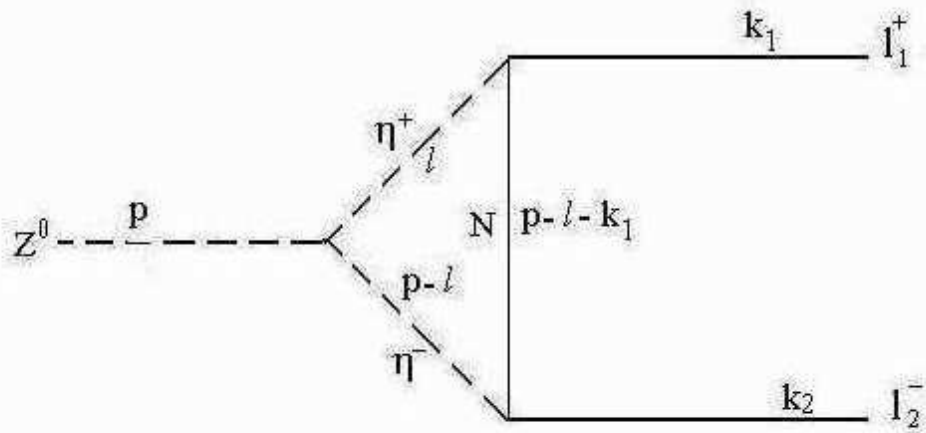
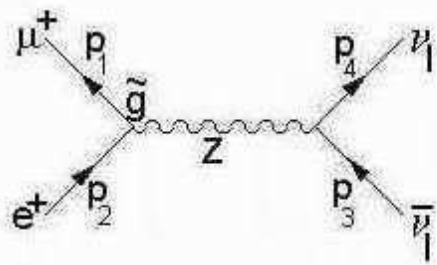
\end{document}